\documentclass[12pt,review,authoryear]{elsarticle}
\usepackage{natbib}
\usepackage{epstopdf}
\journal{Icarus}
\begin{document}
\begin{frontmatter}
\title{Aggregate Particles in the Plumes of Enceladus}
\author[1]{Peter Gao\corref{cor1}}
\ead{pgao@caltech.edu}
\cortext[cor1]{Corresponding author}
\author[1]{Pushkar Kopparla}
\author[2,3]{Xi Zhang}
\author[1]{Andrew P. Ingersoll}
\address[1]{Division of Geological and Planetary Sciences, MC 150-21, California Institute of Technology, Pasadena, CA 91125, USA}
\address[2]{Department of Planetary Sciences and Lunar and Planetary Laboratory, University of Arizona, Tucson, AZ 85721, USA}
\address[3]{Department of Earth and Planetary Sciences, University of California Santa Cruz, Santa Cruz, CA 95064, USA}
\begin{abstract}
Estimates of the total particulate mass of the plumes of Enceladus are important to constrain theories of particle formation and transport at the surface and interior of the satellite. We revisit the calculations of \citet{ingersoll2011total}, who estimated the particulate mass of the Enceladus plumes from strongly forward scattered light in Cassini ISS images. We model the plume as a combination of spherical particles and irregular aggregates resulting from the coagulation of spherical monomers, the latter of which allows for plumes of lower particulate mass. Though a continuum of solutions are permitted by the model, the best fits to the ISS data consist either of low mass plumes composed entirely of small aggregates or high mass plumes composed of mostly spheres. The high particulate mass plumes have total particulate masses of (166 $\pm$ 42) $\times$ 10$^3$ kg, consistent with the results of \citet{ingersoll2011total}. The low particulate mass plumes have masses of (25 $\pm$ 4) $\times$ 10$^3$ kg, leading to a solid to vapor mass ratio of 0.07 $\pm$ 0.01 for the plume. If indeed the plumes are made of such aggregates, then a vapor-based origin for the plume particles cannot be ruled out. Finally, we show that the residence time of the monomers inside the plume vents is sufficiently long for Brownian coagulation to form the aggregates before they are ejected to space. 
\end{abstract}
\end{frontmatter}
\section{Introduction}
Enceladus' plumes provide an indirect way to study the subsurface. In particular, the ratio of ice particle to vapor mass can serve as an important constraint of ice particle formation and transport \citep[henceforth IE11]{ingersoll2011total}. IE11 examined Cassini Imaging Science System (ISS) images of the plumes at small scattering angles from $2.2-5.3^{\circ}$ , where forward scattering is dominated by ice grains as opposed to water vapor. They then fit the resulting phase curves to various mass and shape distributions, assuming that the ice grains are solid spherical or ellipsoidal particles. However, the large particle to vapor mass ratio calculated by IE11 could not be easily explained by any existing theoretical models. Furthermore, recent results from the Cassini Cosmic Dust Analyzer (CDA) (Kempf, S., Cassini Project Science Meeting, Jan 22, 2015, and private communication) indicate that the particulate mass may be a factor of ten lower than the estimates of IE11. Since the defining property of the plume particles is strong forward scattering, an alternative model for the plume is one made up of aggregates. 

An aggregate is a particle of irregular shape composed of smaller subunits (or ``monomers'') stuck to each other. The monomers are usually composed of simpler shapes, such a spheres, plates, or columns, depending on the formation mechanisms. Aggregates form under many different settings. For example, the hazes in the atmospheres of Jupiter and Titan are thought to be composed of hydrocarbon aggregates \citep{west1991evidence,tomasko2008model,zhang2013}; Saturn's F-ring is likely populated by ice aggregate particles \citep{vahidinia2011saturn}; and cirrus cloud particles on Earth take on a variety of non-spherical shapes, ranging from fernlike and fractal geometries to aggregates of irregular shapes \citep{yang1998single}. In particular, ice clouds that form as a result of strong vertical motions are dominated by aggregate particles \citep{baum2011improvements,heymsfield2002general}. Since aggregates form under such a wide variety of conditions, it is plausible that they could form in the plumes of Enceladus.

Thanks to the diversity of instruments on Cassini, the plumes are well studied. There are good estimates of water vapor mass from the Ultraviolet Imaging Spectrograph (UVIS) instrument \citep[for example]{tian2007monte,hansen2011composition}, as well as that of other minor gaseous constituents, such as carbon dioxide, methane, ammonia, and argon from the Ion and Neutral Mass Spectrometer (INMS) \citep{waite2009liquid}. Data from Cassini CDA suggests that the ice particles in the plumes can be broadly classified into two types: slow, large, salt-rich grains that tend to fall back onto the surface, and fast, salt-poor grains that escape into the E-ring \citep{postberg2011salt}. Ice particle velocity distributions were measured using the Visual and Infrared Mapping Spectrometer (VIMS) \citep{hedman2009spectral} and ISS (IE11). However, the shapes of the particles remain mostly unconstrained by observations and are usually assumed to be spherical or oblate/prolate, such as in \citet{porco2006} and IE11. In this study, we derive estimates for the total particulate mass of the Enceladus plumes by extending the range of possible particle shapes to aggregates.

\section{Aggregate Model}
\label{sec:aggmod}

Aggregate particles are defined by two parameters: their fractal dimension $D$ and their monomer radius $r_m$. These two quantities are related to the number of monomers that make up the aggregate $N_m$ and the radius of the aggregate particle $r$ by
\begin{equation}
\label{eq:fractal}
N_m = \left(\frac{r}{r_m}\right)^D.
\end{equation}
\noindent A typical value of $D$ for aggregates in the Solar System is around 2. For example, it has been shown that $D$ = 2 is a good approximation for the aggregate particles in the Titan hydrocarbon haze, where the actual dimension may vary between 1.75 and 2.5 \citep{cabane1993fractal}. $D$ = 2 is also appropriate for snowflakes and cirrus cloud ice crystals on Earth, which have variations in $D$ between 1.9 and 2.3 \citep{westbrook2006,schmitt2010}. \citet{zhang2013} further showed that $D$ = 2 aggregates can be used to fit Cassini ISS observations of the Jupiter stratospheric aerosols. Such particles have masses that scale linearly with surface area, like a sheet, though the particle itself is a three dimensional object. As a result, these particles tend to have small masses associated with large scattering cross sections. 

By comparison, $r_m$ values vary considerably across different types of aggregates. For example, $r_m$ = 10 nm for the stratospheric aerosols of Jupiter \citep{zhang2013} and 40 nm for Titan's haze aggregates \citep{tomasko2009}, but these values are easily dwarfed by that of Saturn's F ring particles, which can reach a few microns \citep{vahidinia2011saturn}, while ice crystal monomers on Earth can be hundreds of microns across \citep{kajikawa1989}. We will therefore leave $r_m$ as a free parameter in our model that will be varied to best fit the data. As a simplification, we assume that all aggregates in the plume have monomers of the same $r_m$. 

\begin{figure}[p]
\centering
\includegraphics[width=1.0 \textwidth, clip=true, trim=0cm 0cm 0cm 0cm]{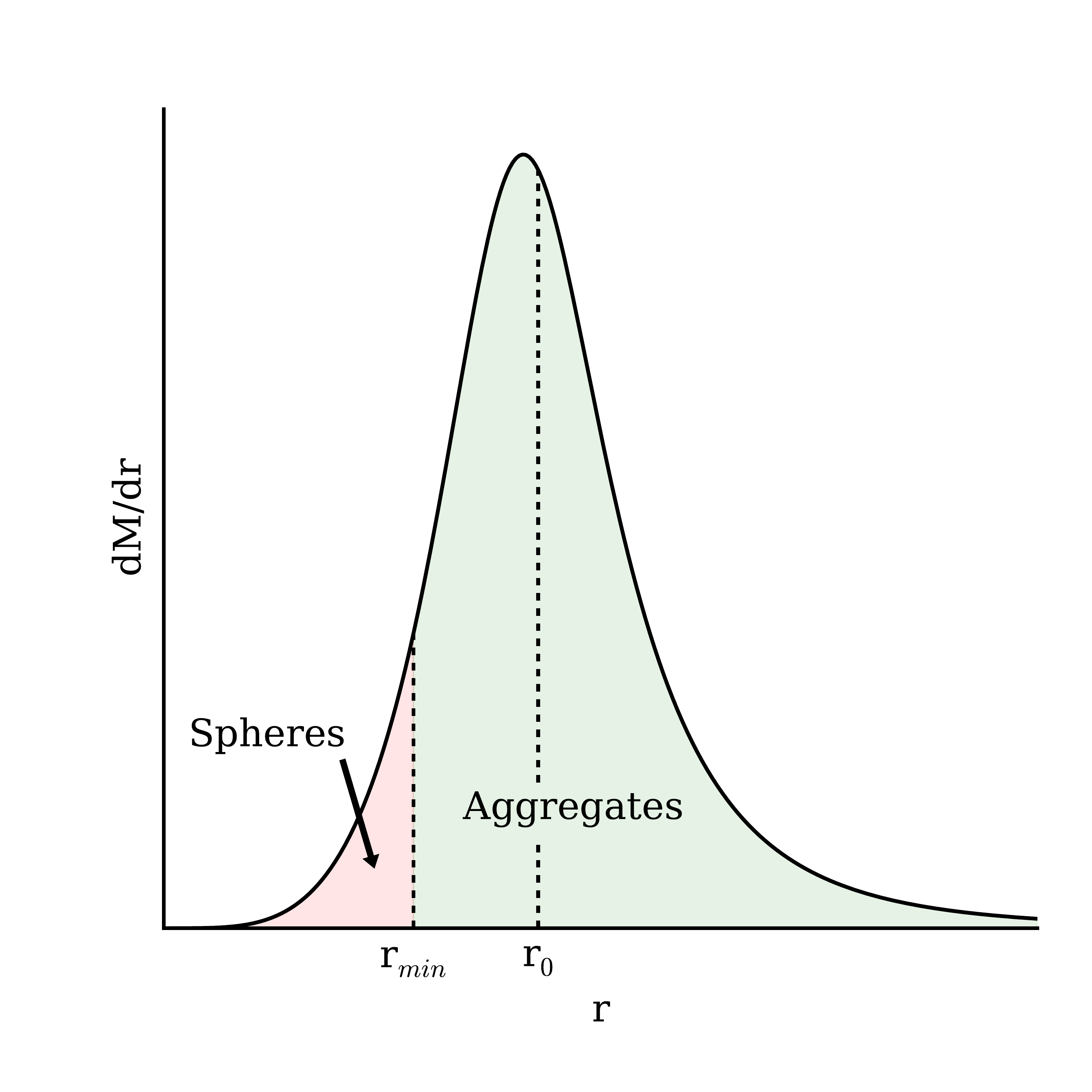}
\caption{A schematic of the plume particle mass distribution $dM/dr$ as a function of the particle radius $r$, with the median particle radius $r_0$ and minimum aggregate radius $r_{min}$ labelled.}
\label{fig:mass_distribution_fig}
\end{figure}

Equation \ref{eq:fractal} leads to a minimum size $r_{min}$ for an aggregate of 
\begin{equation}
\label{eq:rmin}
r_{min} = 2^{\frac{1}{D}} r_m
\end{equation}
\noindent where we have chosen $N_m$ = 2 as the minimum number of monomers an aggregate can have. Particles with $r < r_{min}$ are assumed to be spherical with radius $r$. To simplify the problem and reduce the number of free parameters, we further assume that both spherical and aggregate particles ``share'' the same particle size distribution

\begin{equation}
\label{eq:size_distribution}
\frac{dN}{dlnr} = N_0 (r/r_0)^{f-3}/[1+(r/r_0)^{2f}],
\end{equation}

\noindent which is the number of particles in the natural log of radius interval $d ln(r)$, with $N_0$ as a parameter that scales with the particle number density, $f$ as a positive width factor, and $r_0$ as the median radius of the particle mass distribution given by

\begin{equation}
\label{eq:mass_distribution}
\frac{dM(r)}{dr} = \frac{2M_0}{\pi r_0} \frac{f (r/r_0)^{f-1}}{1+(r/r_0)^{2f}},
\end{equation}

\noindent where

\begin{equation}
\label{eq:mr}
M(r) = \frac{2M_0}{\pi} \arctan \left [ \left ( \frac{r}{r_0} \right)^f \right ]
\end{equation}

\noindent is the total mass of particles with radius between 0 and $r$, and $M_0$ is the total mass of particles. $r_0$ splits the mass distribution into equal halves, with the width of the distribution governed by $f$; small $f$ values indicate wide distributions while large $f$ values indicate narrow distributions. $M_0$, $r_0$, and $f$ are free to vary in the model during optimization of the fit to the data. Eqs. \ref{eq:size_distribution} - \ref{eq:mr} are the same distribution functions given in IE11, which were chosen due to their relative simplicity and their ability to capture both sharply peaked and asymptotic functional forms with only two free parameters. Figure \ref{fig:mass_distribution_fig} shows a schematic of how the aggregate and spherical particles ``share'' the $dM(r)/dr$ distribution in our model. Particles with $r > r_{min}$ are assumed to be aggregates with radius $r$ given in Eq. \ref{eq:fractal} and $r_m$ defined by Eq. \ref{eq:rmin}; they follow the size distribution of Eq. \ref{eq:size_distribution} with some given $r_0$ and $f$ values. Particles with $r < r_{min}$ are assumed to be spheres of radius $r$ that follow the same size distribution as the aggregates, with the same $r_0$ and $f$ values. With these definitions, $dM(r)/dr$ has a discontinuity at $r$ = $r_{min}$. This is caused by the different ways the mass of single particles (aggregate or spheres) scales with $r$ while keeping $N_0$ fixed for both the aggregate and spherical sections of the size distribution. The discontinuity is not shown in Figure \ref{fig:mass_distribution_fig}, and we avoid it in our calculations, as discussed below. 

\section{Observations and Model Setup}
\label{sec:obsms}

Following the procedure of IE11, we use the following relationship to estimate total plume particulate mass (Eq. 4 of that paper):

\begin{equation}
\label{eq:fit_to_data}
R\left(\theta\right) = \frac{M_0}{4\rho _{ice}} \int\frac{A_p Q_{sca} P(\theta)(dN/d ln (r))d ln(r)}{V_p (dN/d ln(r))d ln(r)}
\end{equation}

\noindent where $M_0$ is now the total particulate mass of the plume; $A_p Q_{sca}$ is the scattering cross section of the particle, which, for spherical particles, can be split into the geometric cross section of the particle $A_p$ = $\pi r^2$ and the scattering efficiency  $Q_{sca}$; $P(\theta)$ is the scattering phase function; $\theta$ is the scattering angle, which is given in Table 1 of IE11; $V_p$ is the solid volume of each particle, given by $(4/3) \pi r^3$ for spheres and $(4/3) \pi r_m^3 N_m$ for aggregates; $\rho_{ice}$ = 0.917 g cm$^{-3}$ is the density of ice; and $r$ is the particle radius, as given for aggregates and spheres in Section \ref{sec:aggmod}. $R(\theta)$ is defined as 

\begin{equation}
\label{eq:rtheta}
R\left(\theta\right) = \int \frac{I}{F} dA
\end{equation}

\noindent where $I$ is the measured radiance, $\pi F$ is the solar irradiance, and the integral is taken over the area $A$ of the image projected onto a plane at the distance of Enceladus to Cassini. The values and associated uncertainties of $R(\theta)$ are also listed in Table 1 of IE11. Only the wide-angle camera (WAC) images are used. In order to remove the bias of the data towards scattering angles where more observations were taken, we repeat the procedure of IE11 and grouped the 18 brightness measurements into 10 data points by scattering angle, and averaged the brightness values and uncertainties within each group. 

In order to sample multiple scattering angles the observations were taken at multiple orbital phases. However, \citet{hedman2013observed} showed that the brightness of the plumes varied as a function of orbital phase, even when scattering was taken into account, indicating variability in the intensity of the plumes. They showed that the plume brightness increased rapidly from an orbital phase of 90$^{\circ}$ to 180$^{\circ}$, while below 90$^{\circ}$ the changes in brightness were much more gradual (see their Figure 4). Our observations, as given in Table 1 of IE11, were taken at orbital phases of 7.68$^{\circ}$ to 55.93$^{\circ}$, well within the ``gradual'' part of the orbit. Therefore, while the intensity of the plume still changed during our observational period, we will assume that the changes are minor compared to brightness changes due to particle scattering.

We calculate the Mie scattering cross section and phase function of the spherical particles using \citet{grainger2004}, while those of the aggregates are calculated using the model of \citet[henceforth T08]{tomasko2008model}, a detailed description of which can be found in the appendix of that paper. The T08 model is a fast parametrization valid only for $D$ = 2 due to the computationally intensive calculations required for a model of aggregates of arbitrary dimension \citep{mishchenko1996}; as discussed in Section \ref{sec:aggmod}, $D$ = 2 is valid for many types of aggregates in the Solar System, including ice particles in Earth clouds.  A visual description of such particles is found in Figure 1 of \citet{west1991evidence}. The aggregate scattering model considers each aggregate to be a collection of spherical monomers that scatter and absorb as Mie spheres; the total intensity of the scattered light from the aggregate is then the superposed intensity of scattered light from all the monomers in the aggregate. For more complex interactions, such as repeated scattering and absorption by multiple monomers, correction factors are added that are functions of $N_m$, the refractive index, and the size parameter of the monomers 

\begin{equation}
\label{eq:sizeparameter}
x_m = \frac{2 \pi r_m}{\lambda}
\end{equation}

\noindent where $\lambda$ is the wavelength of the radiation being scattered. The correction factors are empirical and listed in Table A2 of Appendix A of T08. Only cases with $x_m$ between 10$^{-4}$ and 1.5 have currently been validated; and extrapolations to different $x_m$ values become increasingly uncertain the farther they are from the tested range of values (see again Appendix A of T08). The T08 model has been used to fit observations of Titan's haze, as well as that of the stratospheric aerosols of Jupiter \citep{zhang2013} and aggregate particles created in laboratory experiments \citep{bar2008titan}. We assume a real index of refraction of 1.31 and an imaginary index of refraction of 0.0005 for ice in our scattering calculations for the three Cassini ISS wavelength channels considered (VIO, 420 nm; CLR, 635 nm; IR3, 918 nm), as with IE11 \citep{meng2010}. 

In order to avoid the discontinuity in $dM(r)/dr$, we treat particles of different sizes separately by discretizing $dN/dlnr$ into 25 bins in our calculations, with a minimum particle radius of 80 nm and a particle volume ratio of 2 between successive bins. We then approximate the integrals in Eq. \ref{eq:fit_to_data} as summations over the size bins, so that different forms of $A_p$, $Q_{sca}$, $P(\theta)$, and $V_p$ can be used depending on whether the bin corresponds to spherical particles or aggregate particles. This enables each integral to sum over both types of particles across the entire size range, provided $N_0$ is a constant that drops out of Eq. \ref{eq:fit_to_data}, and ensures that $M_0$ is a constant that applies to the entire plume, as $M_0$ = $\rho_{ice} \int V_p (dN/d ln(r))d ln(r)$. The minimum particle size is set to prevent an infinite integral of $dN/dlnr$ for $f$ $\leq$ 3 in our numerical scheme, though the total mass and $R(\theta)$ are always finite for any positive value of $f$. This is because $V_p$ scales as $r^3$ for spheres and $r^2$ for aggregates, and because $Q_{sca}$ scales as $r^4$ when $r \to$ 0. The value of the minimum particle radius was picked such that it is much lower than the likely range of favorable $r_m$ values estimated by fitting the data by hand. We have also confirmed, through tests of our model, that particles smaller than 80 nm do not contribute signficantly to the forward scattering for all reasonable model parameter values.

\section{Fitting to Data}

We use the Markov Chain Monte Carlo (MCMC) method \citep{hastings1970} to optimize the fits to the data and obtain statistically significant confidence intervals for the optimized parameters, which consist of $M_0$,  $r_m$, $r_0$, and $f$, with $D$ fixed to 2. This is the same number of parameters as IE11, though they had a particle aspect ratio rather than $r_m$. 

The MCMC explores parameter space by comparing the likelihood of a model defined by one set of parameters with that defined by a perturbed set of parameters. The likelihood $L$ of a model is defined as
\begin{equation}
\label{eq:llh}
L = e^{\cal L}, {\cal L} = -\frac{1}{2}\sum \limits_{i} \frac{\left(x_i - \mu_i \right)^2}{\sigma_i^2}
\end{equation}
\noindent where $x_i$ is the ith data point, $\mu_i$ is the ith model point, and $\sigma_i^2$ is the variance of the ith data point, calculated from Table 1 of IE11. Models with high $L$ values correspond to better fits to the data than models with low $L$ values, and thus the sets of parameters that define the better-fit models are closer to the true values of these parameters than the sets of parameters that define the worse-fit models. 

At every link of the MCMC, one of the four parameters is chosen randomly to be perturbed by adding onto the original parameter value a perturbation, the magnitude of which is drawn from normal distributions of certain standard deviations: $10^4$ kg for $M_0$; 0.5 $\mu$m for $r_m$ and $r_0$, and 1 for $f$. The likelihood of this perturbed set of parameters is then calculated according to Eq. \ref{eq:llh}. If the likelihood of the model defined by the perturbed set is greater than that of the original set, then the perturbed set of parameters are accepted as a new link in the chain and the new standard with which subsequent sets of perturbed parameters are compared; otherwise, the perturbed set is accepted if the ratio of the likelihood of the model defined by the perturbed set to that of the original set is greater than a random number drawn from a uniform distribution from 0 to 1. In other words, if the $L$ value of the model defined by the perturbed set is very low, i.e. a very bad fit to the data and thus a very low probability that the data is described by this model, then it is highly unlikely that it will be accepted. By calculating the likelihood of models defined by different sets of parameters with a bias proportional to $L$, the MCMC method samples the probability density function (PDF) of the parameters. The PDF of each parameter is then the histogram of the accepted parameter values. 

We use a total of $2 \times 10^6$ links in the MCMC, and the parameter PDFs are constructed from taking the histograms of the accepted values. Sensitivity tests are conducted by halving and doubling the widths of the normal distributions from which the perturbations at each step of the MCMC were drawn, with no major changes to our results. See \ref{sec:appendix} for further details on these sensitivity tests. 

We apply loose bounds to the parameters: $M_0$ is restricted to positive values; $r_0$ is bound between 0.2 and 20 $\mu$m and $r_m$ is limited to values between 0.08 and 14 $\mu$m to ensure that the radius grid captures most of the size distribution; and $f$ is constrained between 0 and 10 to avoid size distributions narrower than the resolution of the radius grid. The best fit parameters of IE11 are well within these bounds. If there arises a perturbation during a step of the MCMC that resulted in values of the parameters outside their bounds, then the perturbation is redrawn from its normal distribution until a value within the bounds results.

\section{Results}

\begin{figure}[p]
\centering
\includegraphics[width=1.0 \textwidth, clip=true, trim=0cm 0cm 0cm 0cm]{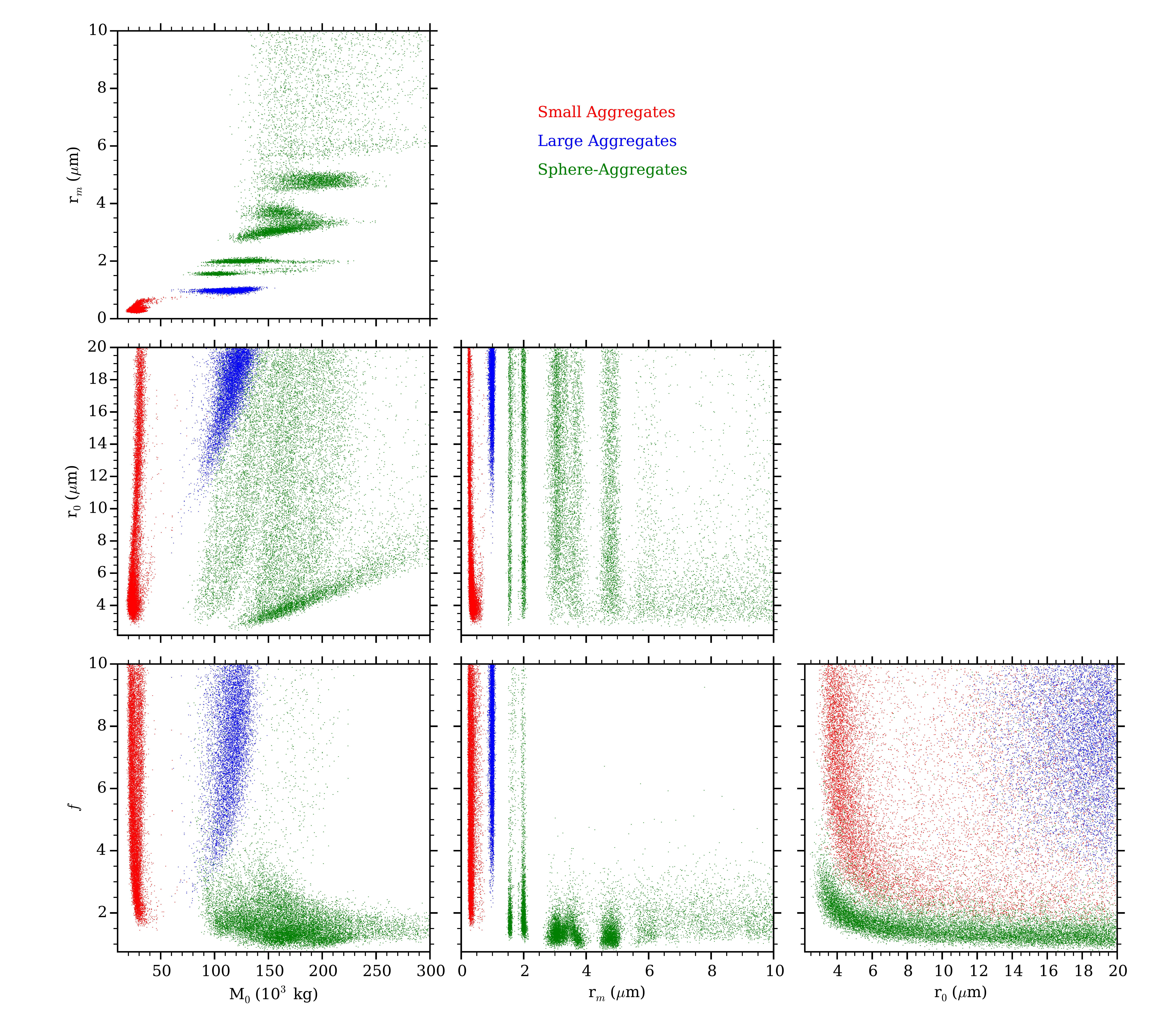}
\caption{Scatter plots of the accepted sets of parameters from the MCMC calculations for each parameter pair. The regions with higher concentrations of points correspond to areas of higher likelihood. Red points correspond to the small aggregate family of solutions defined by $r_m <$ 0.8 $\mu m$, with low $M_0$; blue points correspond to the large aggregate family of solutions defined by 0.8 $\mu m < r_m <$ 1.25 $\mu m$, with high $M_0$, $r_0$, and $f$; and green points correspond to the sphere-aggregate family of solutions defined by $r_m >$ 1.25 $\mu m$, with high $M_0$ and low $r_0$ and $f$.}
\label{fig:mcmcresults_scatter}
\end{figure}

Figure \ref{fig:mcmcresults_scatter} shows the results of the MCMC calculations in the form of scatter plots of the accepted sets of parameters for each parameter pair. Areas of parameter space with more accepted values (points) correspond to places of higher likelihood, and thus better fits to data. It is immediately clear that there are multiple families of solutions, which have been assigned different colored points. For example, the $M_0$ vs. $r_m$ plot shows at least seven local maxima of accepted parameter sets, though the local maximum with the lowest $M_0$ appears to be separate from the others. Conversely, the $M_0$ vs. $r_0$ and $M_0$ vs. $f$ plots show only three local maxima at distinct locations in parameter space, one at low $M_0$ and two at high $M_0$. These two high $M_0$ families of solutions can be further split in terms of $r_0$ and $f$, with one at high $r_0$ and high $f$ (and slightly lower $M_0$) and one at low $r_0$ and low $f$ (and slightly higher $M_0$). From the $r_m$ vs. $r_0$ and $r_m$ vs. $f$ plots we see that the local maximum with the second lowest $r_m$ is resticted to high $r_0$ and high $f$ values, and that it is the only local maximum out of the seven to exhibit this behavior. Therefore, the seven local maxima of $r_m$ can be distributed among the three local maxima in the other plots as follows: the local maximum with $r_m <$ 0.8 $\mu m$ defines the low $M_0$ family of solutions (red); the local maximum with 0.8 $\mu m < r_m <$ 1.25 $\mu m$ defines the high $M_0$, $r_0$, and $f$ family of solutions (blue); and the remaining local maxima with $r_m >$ 1.25 $\mu m$ defines the high $M_0$ and low $r_0$ and $f$ family of solutions (green). In other words, the red points are solutions corresponding to low particulate mass plumes composed primarily of small aggregates with $N_m >$ 100; the blue points are high particulate mass plume solutions consisting of large aggregates made up of large monomers, such that they also have $N_m >$ 100; and the green points are also high particulate mass plume solutions, but they are comprised of a mixture of spheres and large aggregates composed of very few and very large monomers, which we term the ``sphere-aggregate'' solutions. 

Despite the distinctiveness of these solutions in parameter space, most of their x$_m$ values are $>$1.5, which is outside the validated range of the T08 model. Therefore, caution must be exercised when interpreting these results. As there are currently no quantitaive estimates of the divergence of this approximate model from the detailed T-matrix calculations outside the validated range \citep{mishchenko1996}, we compare our results to that of the fractal aggregate scattering model of \citet[henceforth R99]{rannou1999}. The R99 model is semi-empirical and derives particle scattering properties using interpolated values from the mean field approximation of \citet{botet1997}, which itself is based on an exact theory of scattering by aggregate particles \citep{xu1995}. The R99 model has been validated to x$_m$ $\sim$ 8 \citep[for e.g.]{coustenis2001,rannou2002,rannou2004,rannou2010,lavvas2010} and shows good agreement with the T08 model within the latter's validated range. Thus, it can be used as an approximate fiducial to evaluate the accuracy of the T08 model beyond this range. Only rough agreement is necessary however, as the R99 model is more error-prone than the T08 model due to nonlinear effects in light scattering (T08).

\begin{figure}[p]
\centering
\includegraphics[width=0.6 \textwidth, clip=true, trim=0cm 0cm 0cm 0cm]{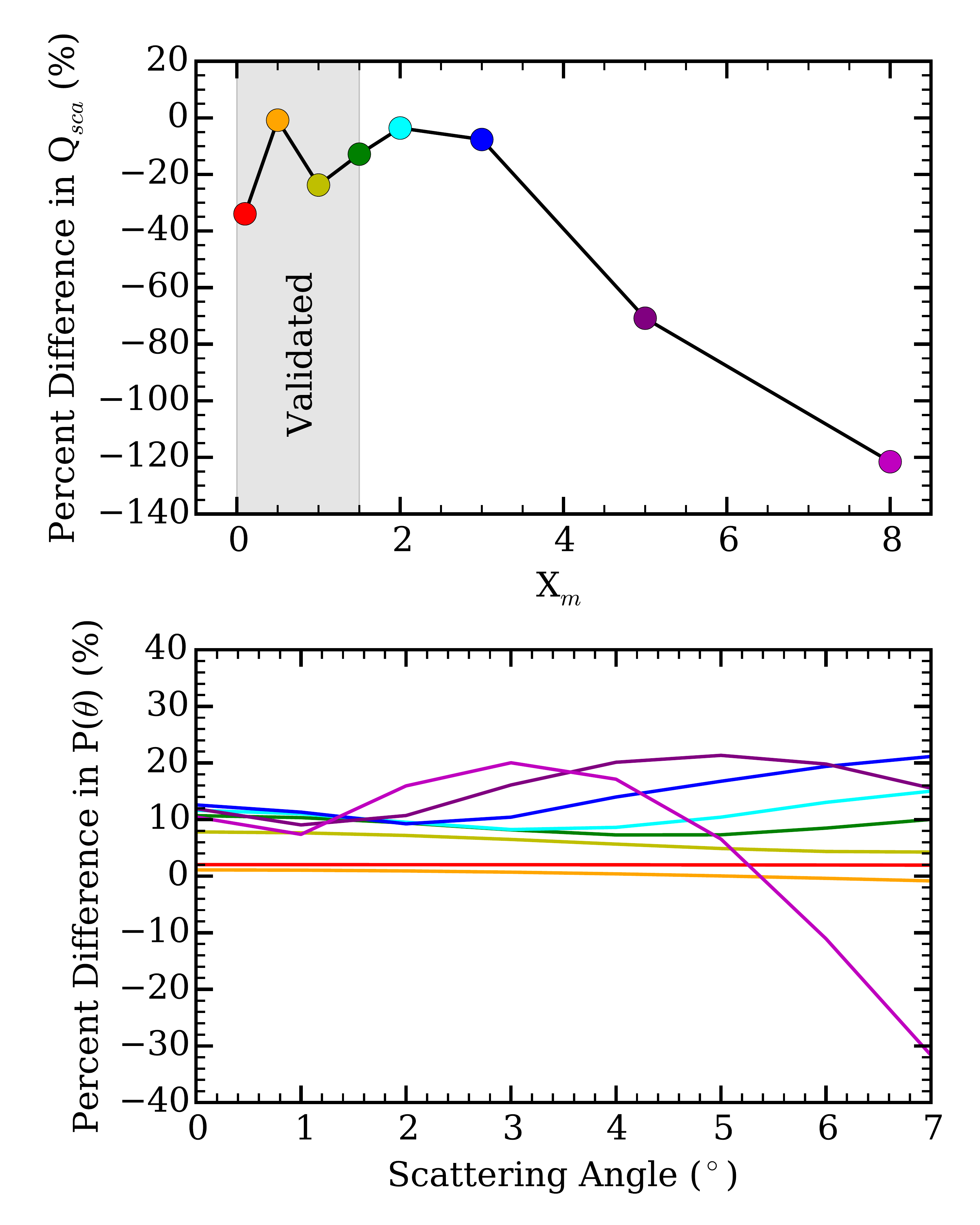}
\caption{Percent difference in the scattering efficiency Q$_{sca}$ (top) and the phase function P($\theta$) (bottom) between the models of \citet{rannou1999} and \citet{tomasko2008model} for relevant values of the size parameter x$_m$ and scattering angle $\theta$. The comparison is done in the CLR wavelength channel with 300 monomers assumed for each case. The refractive indices used are those of water ice (Section \ref{sec:obsms}) and the fractal dimension is set to 2. The x$_m$ values of each of the colored phase curves in the bottom panel are indicated by the same colored points in the top panel. The gray shaded region in the top panel indicates the range in x$_m$for which the model of \citet{tomasko2008model} has been validated.}
\label{fig:rannoutomasko}
\end{figure}

Figure \ref{fig:rannoutomasko} shows the percent difference in the scattering efficiency Q$_{sca}$ (top) and the phase function P($\theta$) (bottom) (see Eq. \ref{eq:fit_to_data}) between the R99 and T08 models for relevant  scattering angles and size parameters. The comparison is done in the CLR wavelength channel with 300 monomers assumed for each case, which is expected for the small and large aggregate solutions given their r$_m$ and r$_0$ values; the refractive indices used are those of water ice (Section \ref{sec:obsms}) and the fractal dimension is set to 2. The x$_m$ values of each of the colored phase curves in the bottom panel are indicated by the same colored points in the top panel. The comparison shows that the Q$_{sca}$ and P($\theta$) values for the two models agree to within 20$\%$ for most x$_m$ $<$ 3.5, thereby extending the range of validity of the T08 model. For the small aggregates, x$_m$ $\sim$ 3 for the CLR channel (assuming r$_m$ $\sim$ 0.3 $\mu$m), and thus it is a valid family of solutions. Conversely, the large aggregates have x$_m$ $\sim$ 10 for the CLR channel (assuming r$_m$ $\sim$ 1 $\mu$m), which is well outside the extended validity range, and therefore calls into question the existence of this family of solutions. An additional issue with the large aggregate solutions is its high r$_0$ value ($\sim$ 20 $\mu$m), which goes against previous estimates of the sizes of the Enceladus plume particles of a few $\mu$m \citep{kieffer2009,hedman2009spectral}. The sphere-aggregate solutions are largely independent of the limitations of the T08 model, as most of the forward scattering is due to spherical particles.

From this exercise it is clear that the small and sphere-aggregate solutions are likely real, but that the large aggregate solutions may not be. Therefore, we will ignore the large aggregates in the remainder of this work.

\begin{figure}[p]
\centering
\includegraphics[width=0.6 \textwidth, clip=true, trim=0cm 0cm 0cm 0cm]{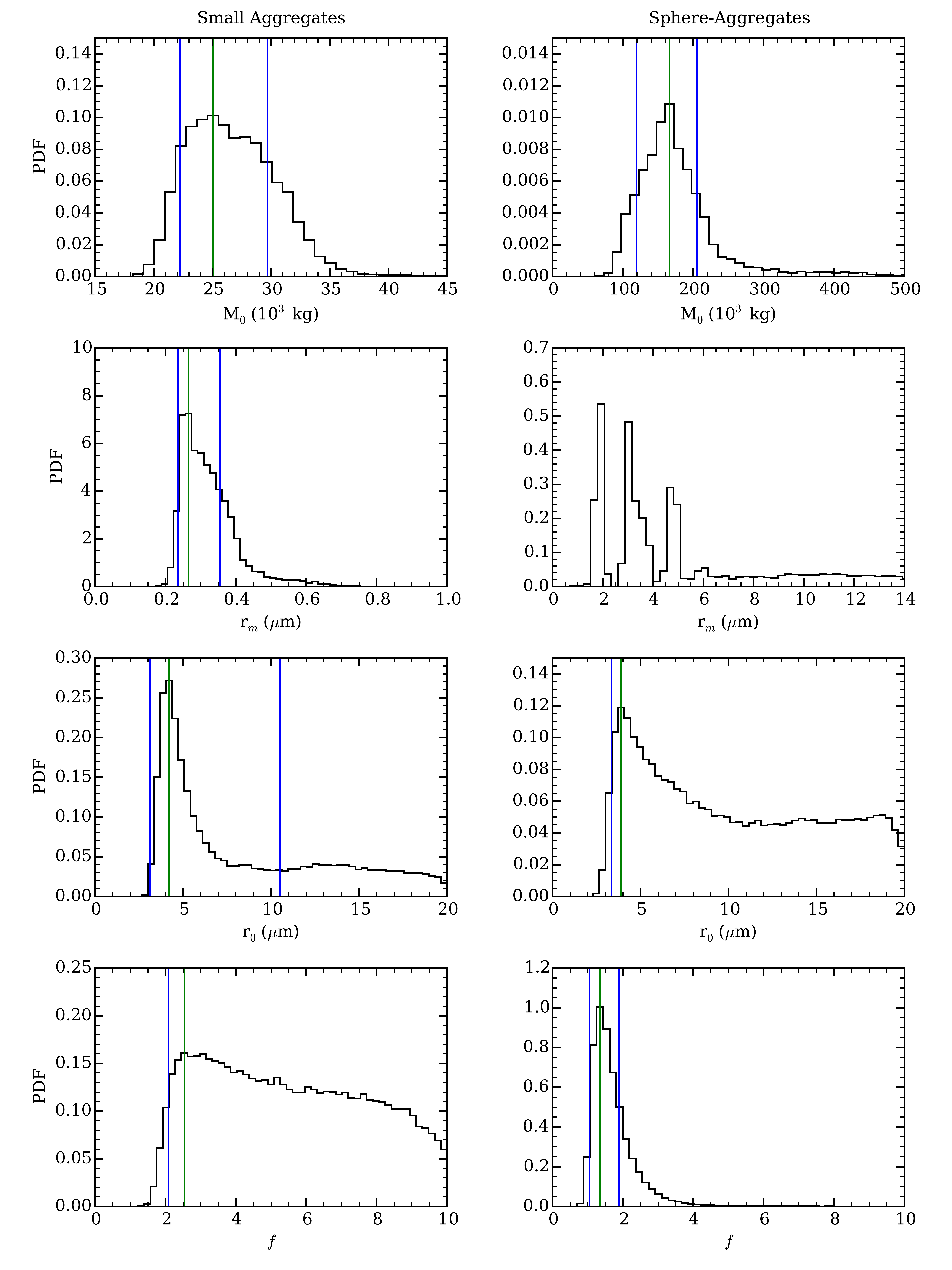}
\caption{The probability density functions (PDFs) of the total mass of the plume $M_{0}$ (top), the monomer radius $r_{m}$ (middle-top), the median particle radius $r_{0}$ (middle-bottom), and the width of the size distribution $f$ (bottom) for the small aggregate (left) and sphere-aggregate (right) particle plume solutions. Note the different abscissa values between the $r_0$ and $r_m$ plots. Each histogram contains 50 bins. The most probable value is marked by the green line. For the $M_0$, small aggregate $r_m$ and $r_0$, and sphere-aggregate $f$ PDF's, the 68$\%$ confidence intervals are enclosed by the blue lines, while for the small aggregate $f$ and sphere-aggregate $r_0$ PDFs only the lower 68$\%$ confidence interval is marked. The sphere-aggregate $r_m$ PDF is multimodal and thus does not have a well-defined most-probable value or confidence interval.}
\label{fig:mcmcresults_hist}
\end{figure}

The left side of Figure \ref{fig:mcmcresults_hist} shows the PDFs of the four varied parameters for the small aggregate solution, generated by marginalizing the 2D distributions of Figure \ref{fig:mcmcresults_scatter} along one dimension for the small aggregate solution only and normalizing the area of the resulting 1D histograms. The most probable values - the maximum likelihood estimates - correspond to the peaks of the PDFs, marked by the green lines. Models generated from these parameter values are most likely to describe the data. For the $M_0$, $r_m$, and $r_0$ PDFs, the 68$\%$ (1$\sigma$) confidence intervals are enclosed by the blue lines, which indicate the smallest interval in parameter space that encloses 68$\%$ of the accepted values. The PDF of $f$ does not have a well-defined peak, and thus a lower limit is given, with the 68$\%$ confidence interval now defined as the 68$\%$ of accepted values immediately below the most probable value. Table \ref{tab:bestfitparams} gives these most probable values and the associated 68$\%$ confidence intervals, which we set to be the ranges of uncertainty in our fit to data. 

The small aggregate plume solution yields a most probable $r_0$ consistent with the $r_0$ of the spherical particles of IE11, but the total plume mass is $\sim$ 6 times lower. Meanwhile, the PDF of $f$ is very broad, meaning that this parameter is not well-constrained by the Cassini ISS data. The lack of a upper limit allows for large $f$ values, leading to narrow size distributions centered at $r_0$, with low abundances of both smaller and larger particles. This is consistent with a scenario where the smaller monomers are depleted to form small aggregates of median radius $r_0$.

The right side of Figure \ref{fig:mcmcresults_hist} shows the PDFs of the sphere-aggregate solutions. As before, the 68$\%$ confidence intervals are shown for PDFs with well-defined peaks ($M_0$ and $f$), while for the $r_0$ PDF only the lower limits are given. The $r_m$ PDF is multimodal and thus does not have a well-defined most-probable value or confidence interval. This is likely due to the limitations of the T08 model, as the corresponding x$_m$ values are far greater than even the extended validity range given above. Beyond r$_m$ $>$ 6 $\mu$m, spherical particles dominate the particle population, and thus $r_m$ becomes unconstrained. 

The results of IE11 indicate that, for spherical particles, mass distributions with $f$ = 1 to 2 best fit the data. This is consistent with our sphere-aggregate particle plume results. Our $M_0$ and $r_0$ estimates for the sphere-aggregate solutions are also consistent with IE11, though the PDF for $r_0$ has a significant tail towards larger values. This highlights a deficiency in forward scattering measurements, as more massive particles are difficult to observe due to their low surface to volume ratio. Therefore, the impact of massive particles on the phase curves is minimal, allowing the results to be poorly constrained at large particle sizes. 

Figure \ref{fig:bestfits} gives representative best fits to the VIO (top), CLR (middle), and IR3 (bottom) channels of the Cassini ISS data for the small aggregate (red) and sphere-aggregate (green) particle solutions, the latter of which reproduces the lower left panel of Figure 6 of IE11. In our case, the small aggregate solution provides the best fit, as defined by its reduced chi-square value

\begin{equation}
\label{eq:redchisqrt}
\chi^2_r = \frac{1}{N-n-1}\sum \limits_{i} \frac{\left(x_i - \mu_i \right)^2}{\sigma_i^2}
\end{equation}

\noindent where $N$ = 10 is the number of data points being fitted to, $n$ = 4 is the number of parameters, and the variables in the summation are the same as that of Eq. \ref{eq:llh}. Lower $\chi^2_r$ values indicate better fits, though if $\chi^2_r \ll 1$ then the error bars on the observations are likely overestimated. $\chi^2_r$ = 0.71 for the small aggregate solution fit and $\chi^2_r$ = 1.5 for the sphere-aggregate solution fit. Both solutions give similar fits for the VIO and CLR channels, but differ significantly in the IR3 channel at scattering angles $< 2^{\circ}$. Whereas the small aggregate model curve features a broad forward scattering peak, the sphere-aggregate model curve's forward scattering peak is much narrower in IR3. The width of the forward scattering peak is related to $r_0$ (IE11), but also $r_m$ in the case of aggregates \citep{lavvas2010}; large particles result in narrower/sharper peaks while smaller particles result in broader peaks. Therefore, as the small aggregate particles have small $r_m$ values, their forward scattering peaks are the most broad, increasing $R(\theta)$ at larger scattering angles and prompting low $M_0$ values. Meanwhile, the lack of small monomers in the sphere-aggregate particles narrows the forward scattering peak, resulting in low values at large $\theta$, and a higher $M_0$ than for the small aggregate solution. 

It is interesting to note that only the IR3 channel offers a way to discriminate between the two solutions. This is caused by the strong dependence of the intensity of scattered light on the particle size parameter $x_m$. In our fits there are two ranges of particle sizes to consider: the small aggregate monomer radius $\sim$ 0.3 $\mu$ m and the median particle radius for the small aggregate and sphere-aggregate solutions $\sim$ 4 $\mu$m; the intensity of the scattered light will depend on a combination of the particle radius and the monomer radius, if applicable. For the VIO and CLR channels ($\lambda$ = 420 and 635 nm, respectively), both the small aggregate and sphere-aggregate particles are much larger than the wavelengths considered, leading to narrow forward scattering peaks. However, for the IR3 channel ($\lambda$ = 918 nm), the small aggregate monomer radius is now much smaller than the wavelength, leading to a much broader forward scattering peak. Similarly, the low $f$ value of the sphere-aggregate solution means that there exists a great number of small particles that are responsible for much of the scattering, which are now much smaller than the channel wavelength, again leading to a wider forward scattering peak. 

In obtaining two separate families of solutions, we have shown that degeneracy exists in the ISS data from IE11 when both spherical and aggregate particles are considered. This degeneracy can be broken by an independent mass measurement, such as from Cassini CDA. 

\section{Discussion}
\label{sec:discussion}

\begin{figure}[p]
\centering
\includegraphics[width=0.6 \textwidth, clip=true, trim=0cm 0cm 0cm 0cm]{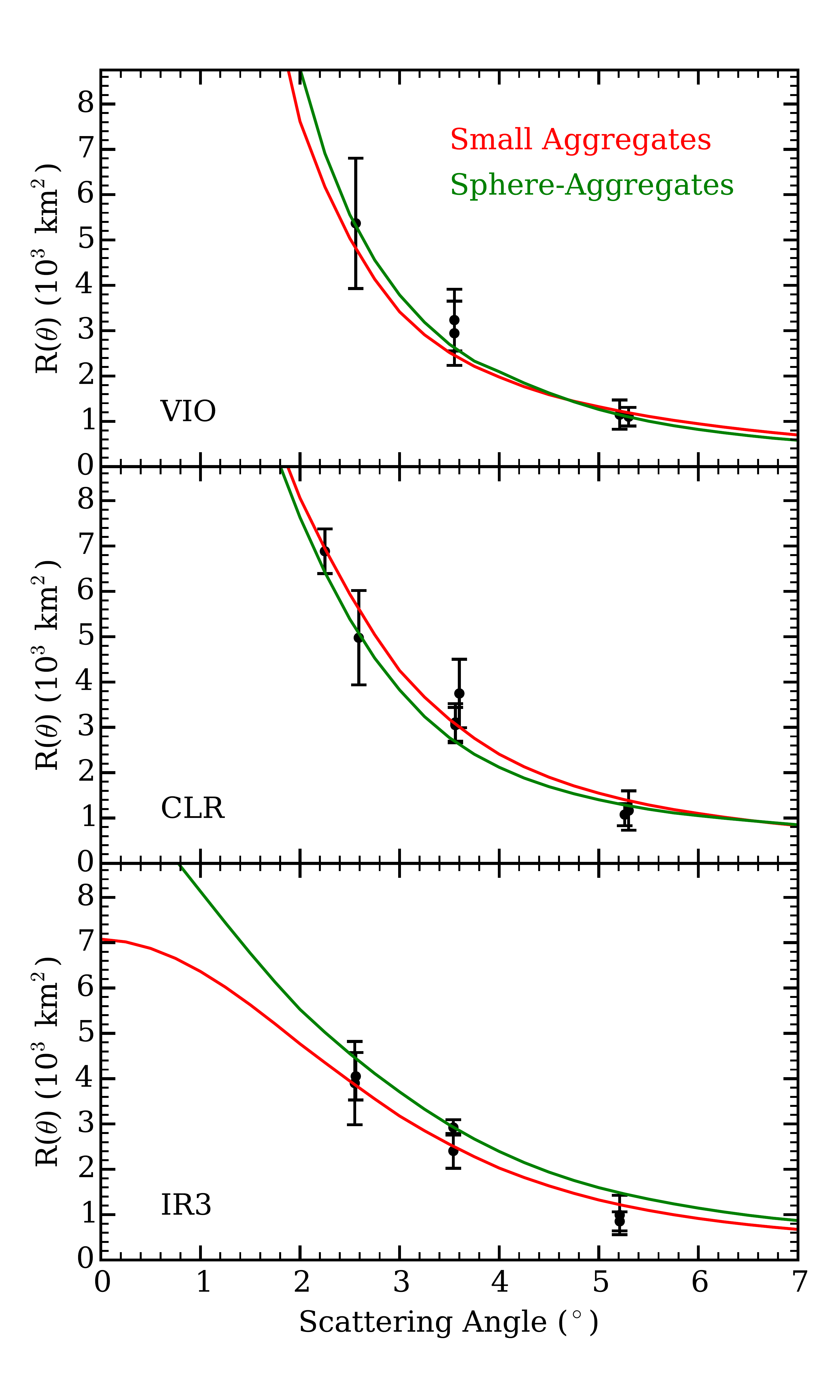}
\caption{Representative best fits to the VIO (top), CLR (middle), and IR3 (bottom) wavelength channel data from Cassini ISS for the small aggregate (red) and sphere-aggregate (green) plume particle solutions. Parameters used for the small aggregate solution fit are: $M_0$ = 22.58 $\times$ 10$^3$ kg, $r_m$ = 0.331 $\mu$m, $r_0$ = 3.9 $\mu$m, and $f$ = 7.79. Parameters used for the sphere-aggregate solution fit are: $M_{0}$ = 172.42 $\times$ 10$^{3}$ kg, $r_{m}$ = 4.87 $\mu$m, $r_{0}$ = 4.81 $\mu$m, and $f$ = 1.71.}
\label{fig:bestfits}
\end{figure}

We can calculate the solid to vapor mass ratio of the plume for the small aggregate solution following the procedure described in IE11. We derive a value for the particulate column mass abundance $M^p_{col}$ by relating it to the $I/F$ values of the Cassini ISS NAC image from IE11 using their Eq. 8,

\begin{equation}
\label{eq:ie11eq8}
\frac{I}{F} =  \frac{M^p_{col}}{M_0}R(\theta) = K_0(\theta)M^p_{col},
\end{equation}

\noindent where $R(\theta)$ is defined in Eq. \ref{eq:rtheta}. Dividing $M^p_{col}$ by the vapor column mass abundance, $M^v_{col}$, at the same altitude above Enceladus then gives the solid to vapor mass ratio of the plume at that altitude. From the small aggregate curve (red) in Figure \ref{fig:bestfits}, we interpolate a $R(\theta)$ value of 6.4 $\times$ 10$^3$ km$^2$ for the CLR channel at $\theta$ = $2.38^{\circ}$, the wavelength channel and scattering angle of the NAC image, respectively. This is slightly higher than the value given in IE11, 5.4 $\times$ 10$^3$ km$^2$, owing to the different model curves used for the interpolation. Setting $M_0$ to (25 $\pm$ 4) $\times$ 10$^3$ kg for the small aggregate solution and $I/F$ to 0.07, the brightest pixel value of the NAC image above background (as with IE11), we find that $M^p_{col}$ = (2.7 $\pm$ 0.4) $\times$ 10$^{-7}$ kg m$^{-2}$ for that pixel, which is $\sim$ 7 times lower than the $M^p_{col}$ value calculated by IE11. The altitude above Enceladus that corresponds to that pixel is similar to that probed by the occultations carried out by \citet{hansen2011composition} to measure  $M^v_{col}$ using Cassini UVIS. Therefore, the solid to vapor mass ratio can be obtained by dividing the $M^p_{col}$ value derived here by the $M^v_{col}$ value derived from averaging the second column of Table 1 of  \citet{hansen2011composition}. This gives a solid to vapor mass ratio for the small aggregate plumes of

\begin{equation}
\label{eq:masstovapratio}
\frac{M^p_{col}}{M^v_{col}} = 0.07 \pm 0.01 ,  
\end{equation}

\noindent which is again 7 times lower than that of IE11. The error arises from uncertainties in both the solid and vapor column mass abundances and is likely underestimated, as we have assumed zero uncertainty for the $R(\theta = 2.38^{\circ})$ and $I/F$ values. It should be noted that, while $K_0(\theta)$ is defined using all available WAC images, the comparison with UVIS requires that we use just the brightest pixel of the one NAC image to set the $I/F$ value. In other words, the global $I/F$ values of the plume in the WAC images are assumed to be the same as that of the brightest plume pixel in the NAC image, which is equivalent to assuming that the particle size distribution is independent of altitude above Enceladus. This is only an approximation, as evidence exists that the particle size distribution does change with altitude \citep{hedman2009spectral,postberg2011salt}. However, for simplicity we do not consider this effect in this work.

Our calculated solid to vapor ratio is consistent with that of \citet{kieffer2009}, who provided an upper bound of $\sim 0.1-0.2$ using observations from Cassini ISS and UVIS \citep{porco2006}. It is also consistent with the lower end of the range of solid to vapor ratios ($\sim 0.01-100$) calculated by \citet{hedman2009spectral} using Cassini VIMS observations and spherical and irregular particle scattering models, though aggregates were not considered.

\begin{figure}[p]
\centering
\includegraphics[width=1.0 \textwidth, clip=true, trim=0cm 0cm 0cm 0cm]{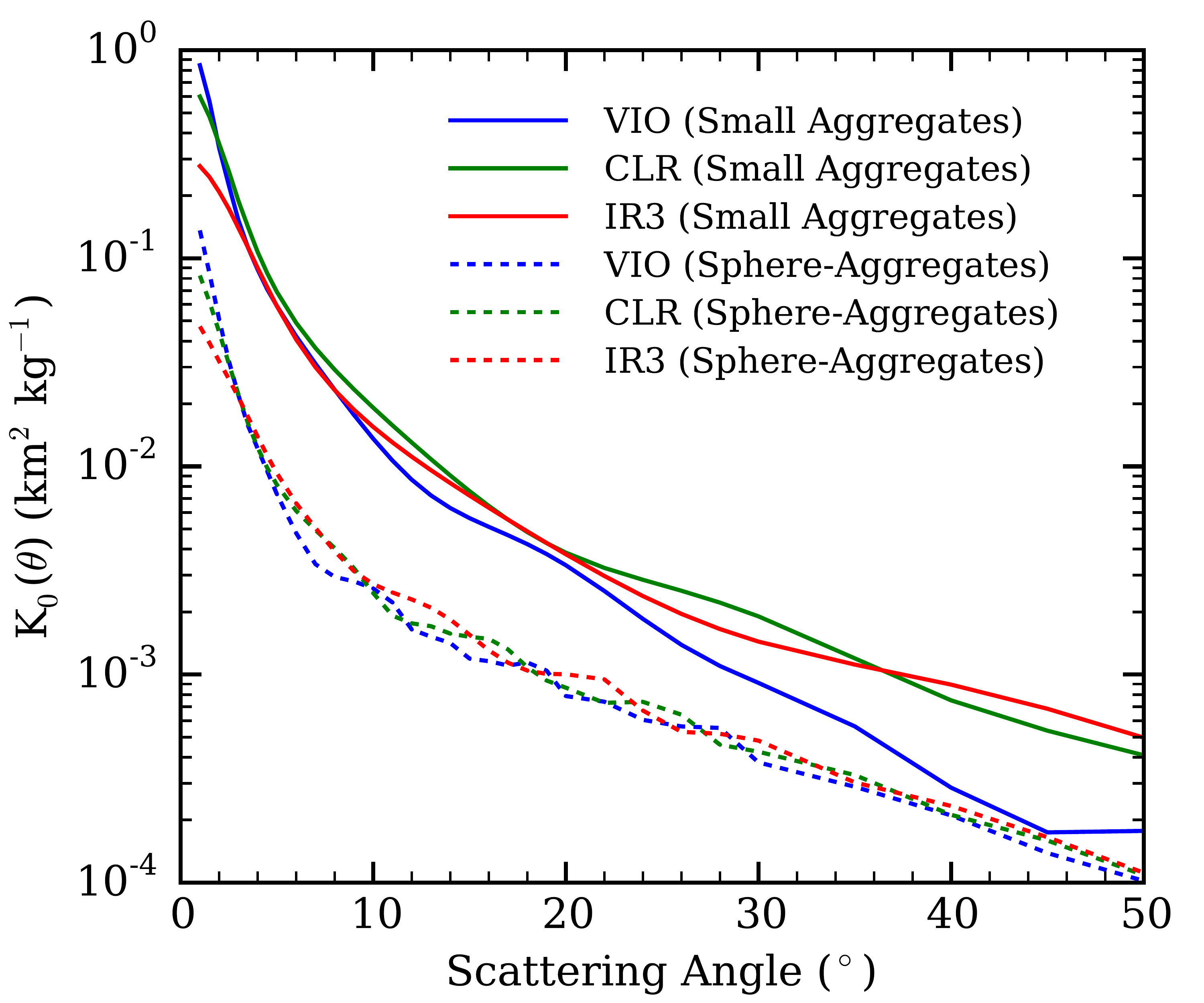}
\caption{Variations of K$_0$($\theta$) = R($\theta$)/M$_0$ as a function of scattering angle for the VIO (blue), CLR (green), and IR3 (red) wavelength channels and the small aggregate (solid lines) and sphere-aggregate (dashed lines) solutions.}
\label{fig:k0curves}
\end{figure}

The $K_0$($\theta$) quantity in Eq. \ref{eq:ie11eq8} is extremely useful as it can readily convert an observed $I/F$ value into the column mass abundance. It also varies depending on the particle scattering properties, wavelength of scattered light, and particle size distribution. Figure \ref{fig:k0curves} shows the K$_0$($\theta$) values of the small (solid lines) and sphere-aggregate (dashed lines) solutions at scattering angles between 0$^{\circ}$ and 50$^{\circ}$ for the VIO (blue), CLR (green), and IR3 (red) wavelength channels. The small aggregate K$_0$($\theta$) values are, as expected, higher than that of the sphere-aggregate solution, since there is less mass for the same intensity of forward scattering. The sphere-aggregate K$_0$($\theta$) values are typically within 10$\%$ of those of IE11, further reinforcing the lack of dependence of the sphere-aggregate solutions on the limitations of the T08 model. A full set of K$_0$($\theta$) values for the two families of solutions and the best fit solution of IE11 at the Cassini WAC and NAC channel wavelengths can be found in the online supplemental material.

Aside from what can be retrieved from observations, it is also imperative that the particle solutions are physical and adhere to theoretical constraints of formation and evolution. As IE11 has already considered high mass plumes, we will focus on the low mass small aggregate solutions and their possible formation mechanisms for the remainder of this work.

Aggregate particles in the Solar System typically form through the coagulation of monomers, a process that depends heavily on the free-floating monomer number density \citep[for e.g.,]{lavvas2010}, i.e. those not already incorporated into aggregates. Therefore, much of the coagulation process must take place within the plume vents, where the particle number density is higher. This is consistent with our small aggregate particle solution, which shows a low abundance of monomers compared to aggregates in the plume, as the monomers must have already coagulated to form aggregates before they were ejected into space. The residence time of particles inside the plume vents is of order $D/v$, where $D$ is the depth of the liquid vapor interface and $v$ is the average velocity of the particles. In order for complete coagulation of monomers within the plume vents, this residence time must be greater or equal to the coagulation time scale, approximated by 

\begin{equation}
\label{eq:coagtime}
\tau_{coag} = \left(\frac{1}{n} \left | \frac{dn}{dt} \right | \right)^{-1} = \frac{1}{nK},
\end{equation}

\noindent where $n$ is the free-floating monomer number density in the vent, and $K$ is the coagulation kernel. In the free molecular limit (Knudsen number $>>$ 1), the Brownian coagulation kernel for two spherical particles with radii $r_1$ and $r_2$ is

\begin{equation}
\label{eq:coagkernelfreemol}
K = \left(r_1+r_2\right)^2 \sqrt{\frac{6 k T}{\rho_{p}}\left(\frac{1}{r_1^3}+\frac{1}{r_2^3}\right)},
\end{equation}

\noindent where $k$ is the Boltzmann constant, $T$ is temperature, and $\rho_{p}$ is the solid particle density (0.917 g cm$^{-3}$). Setting $r_1$ = $r_2$ =  $r_m$, 

\begin{equation}
\label{eq:coagkernelfreemolsimp}
K = \sqrt{\frac{192 k T r_m}{\rho_{p}}}
\end{equation}

\noindent \citep{pruppacher1978}. To find a value for the coagulation time scale, we consider the rate at which particulate mass is emerging from the vents, which can be expressed as 

\begin{equation}
\label{eq:massrate}
\frac{dM_{p}}{dt} = \frac{4}{3} \pi r_m^3 \rho_p n A v=\rho_m A v, 
\end{equation}

\noindent where $\rho_m$ is the total mass of particulates per unit volume and $A$ is the total vent area of the Enceladus plumes. IE11 gives this rate as $dM_{p}/dt$ = 51 kg s$^{-1}$, but the reduced plume mass of the small aggregate solutions results in a smaller value. Rather than assuming that $dM_{p}/dt$ is linearly proportional to $M_0$ however, we instead appeal to the solid to vapor mass ratio by considering the rate at which vapor is emerging from the vents, $dM_{v}/dt$, which has been determined by \citet{hansen2011composition} to be $\sim$ 200 kg s$^{-1}$ based on Cassini UVIS observations. Thus,

\begin{equation}
\label{eq:vapormassrate}
\frac{dM_{v}}{dt} = \rho_v A v = 200\ \textrm{kg}\ \textrm{s}^{-1}, 
\end{equation}

\noindent where $\rho_v$ is the density of the vapor at the liquid-vapor interface, 4.85 $\times$ 10$^{-3}$ kg m$^{-3}$ at T = 273 K. Dividing Eq. \ref{eq:massrate} by Eq. \ref{eq:vapormassrate} gives us the solid to vapor mass ratio $\sim$ [$dM_{p}/dt$ / $dM_{v}/dt$] and allows us to eliminate $A$ and $v$. Substituting the result into Eq. \ref{eq:coagtime} then allows us to eliminate $n$. The necessary condition for the completion of monomer coagulation before exiting the vent then becomes

\begin{equation}
\label{eq:coagulationcondition}
\frac{D}{v} > \tau_{coag} = \frac{1}{K} \left (\frac{4}{3} \pi r_m^3 \right ) \frac{\rho_p}{\rho_v} \frac{dM_v/dt}{dM_p/dt} \sim 20\ \textrm{s}
\end{equation}

\noindent where we use $r_m$ = 0.3 $\mu$m and $T$ = 273 K. For [$dM_{p}/dt$ / $dM_{v}/dt$] we use the solid to vapor mass ratio derived in Eq. \ref{eq:masstovapratio}.

Several assumptions have gone into Eq. \ref{eq:coagulationcondition}. One is that the particles and the vapor emanate from the same liquid-vapor interface, or at least very close above it in the vent. This is necessary to maintain high $T$ and $\rho_p$ values. If the particles form in flight higher up in the vent, as in a convective cloud, the reduction in $\rho_p$ due to lower $T$ and/or condensation on the vent walls would drastically increase $\tau_{coag}$. The fact that some of the plume particles are Na-rich \citep{postberg2009,postberg2011salt} argues that they originate from the liquid, but we cannot rule out a vapor origin for the Na-poor particles. Another assumption is that the different parts of the erupting area $A$ behave the same way whether they are a single crack running the length of each tiger stripe, a series of 100 discrete sources \citep{porco2014}, or some combination of the two \citep[for e.g.]{tian2007monte,hansen2011composition,postberg2011salt,spitale2015}. A third assumption is that the solid to vapor mass ratio remains the same as the plumes vary with orbitial phase \citep{hedman2013observed}. These assumptions are difficult to test, but they seem reasonable. 

A fourth assumption is that the relevant vapor density and velocity are close to those at the liquid-vapor interface and not those at the top of the vent where the particles and vapor exit to space. From the vertical distribution of particles exiting the vent, one infers velocities of order 60\textendash 90 m s$^{-1}$ (Porco et al., 2006; IE11). However, the particles are accelerated to this speed within a few meters of the surface \citep{schmidt2008,ingersoll2010}, so the exit speed underestimates the time spent in the vent during which coagulation takes place. \citet{postberg2009} use 500 m s$^{-1}$ for the speed of the vapor and the triple point value for its density, but \citet{nakajima2015} point out that these are overestimates. Their model includes friction with the walls of the crack as the gas flows upward. This produces backpressure at the liquid-vapor interface, and for a long, narrow crack the backpressure severely limits the evaporation rate and the upward velocity. For crack widths less than 0.1 m, the upward velocity is less than 10 m s$^{-1}$ below depths of 1 km. If the evaporating surface is another 1 km below that level, the time spent in the vent is well over 100 s and the condition for coagulation of aggregates (Eq. \ref{eq:coagulationcondition}) is comfortably satisfied. Crack width is an important parameter, since it affects the velocity. \citet{nakajima2015} derive their estimate by matching the ratio of the power released through latent heat to that released through infrared radiation, but one should remember that these numbers are uncertain by factors of 2 or more. 

Our results favor a less violent source for the plumes than the earlier results of IE11. Here, ``less violent'' means steady, controlled evaporation rather than explosive boiling. For example, it could be that most of the ice particles condense directly from the expanding vapor above the liquid-vapor interface rather than erupt as a bubbly liquid, which then breaks up and freezes into small ice particles in the vacuum of space. Alternatively, the particles could be generated by bubbles of methane or CO$_2$ rising through liquid water and breaking at the surface, thereby sending up a fine spray of ice particles. \citet{porco2006} mentioned two extremes: particles condensing directly from the vapor and particles forming from the breakup of a boiling liquid, the latter analogous to a cold Yellowstone geyser. The relatively large solid to vapor ratio of IE11 favors the violent source, while the lower ratio of our small aggregate solution allows for less violent sources. Thus, the present results make the ISS data compatible with a wider variety of sources than the IE11 results.

\citet{schmidt2008} and \citet{ingersoll2010} used hydrodynamic models to simulate the condensation of ice particles from water vapor, and obtained solid to vapor mass ratios of 0.05 - 0.06 and 0.015, respectively. The lower value calculated by \citet{ingersoll2010} results from the condensation of vapor onto the vent walls, which \citet{schmidt2008} did not consider. These values of the solid to vapor mass ratios are more consistent with our small aggregate solution than with the results of IE11. \citet{kieffer2006} considered explosive decomposition of a hydrate clathrate that contains methane and other gases in much greater abundances than their solubility in liquid water would allow. The solid to vapor mass ratio of such an event is uncertain, but it could be lower than that quoted by IE11 and closer to the ratio derived here from the small aggregate solution. Meanwhile, \citet{hsu2015} and \citet{postberg2009,postberg2011salt} showed that silica grains and sodium salts are present in both E ring and plume particles, which argues strongly for frozen droplets originating from a salty liquid rather than ice particles forming directly from vapor condensation. \citet{postberg2009} suggested that ascending bubbles of plume gases (CO$_2$, N$_2$, CO, CH$_4$) can disperse the liquid droplets into the vapor, and that the evaporation of the boiling liquid source is taking place over a large horizontal area so that freezing of the vents is suppressed. \citet{nakajima2015} showed that the backpressure due to friction on the walls of the channel could severely limit the evaporation rate. \citet{ingersoll2015} argue that this ``controlled boiling'' allows bubbles of vapor to break at the surface instead of throwing up large amounts of spray as in boiling into vacuum. All of this points to a relatively gentle source and a relatively low solid to vapor mass ratio. 

Our small aggregate solution does not rule out a vapor-based source for the plume particles. \citet{parkinson2008} showed that spherical particles with radii similar to $r_m$ can form from water vapor nucleating on involatile grains in a few seconds, while spherical particles with radii up to 3 $\mu m$ can form by condensation in about 100 s, though large condensation nuclei and/or a high density of water vapor may be necessary. Similarly, non-spherical ice crystals resembling aggregates, e.g. dendritic shapes, can form directly from condensation and nucleation of water vapor, with growth rates comparable to that of spherical particles \citep{pruppacher1978}. Bullet rosettes -- highly non-spherical, polycrystalline ice particles made up of clumps of hexagonal columns -- can also yield high scattering cross sections for a small particle mass, but they are formed by the fracturing of ice during the freezing of a large water drop \citep{baum2011improvements}.

We have shown that forward scattering observations of the Enceladus plumes from Cassini ISS can be well-fit by plumes made up of ice aggregates and solid ice spheres. This results in a bifurcation in the allowed total particulate mass of the plumes: small aggregate plumes are six times less massive than plumes made of larger aggregates and spheres. The small aggregate plumes, in particular, lead to a solid to vapor mass ratio of 0.07 $\pm$ 0.01, which is suggestive of a ``gentle source'' for the plume particles. 

\bigskip

\noindent We thank Y. L. Yung and H. Ngo for their valuable input. P. Gao and P. Kopparla were supported in part by an NAI Virtual Planetary Laboratory grant from the University of Washington to the Jet Propulsion Laboratory and California Institute of Technology under solicitation NNH12ZDA002C and Cooperative Agreement Number NNA13AA93A. X. Zhang was supported by the Bisgrove Scholar Program at the University of Arizona. A. P. Ingersoll was supported by the Cassini Project and NASA's Cassini Data Analysis Program.

\bibliographystyle{elsarticle-harv}
\bibliography{encrefs}

\clearpage


\clearpage

\clearpage

\clearpage

\clearpage


\clearpage

\begin{table}[p]
\centering
\begin{tabular}{l c c}
\hline
Parameter & Small Aggregates  & Sphere-Aggregates \\
\hline
$M_0$ (10$^3$ kg) 	& 25 (22 - 30) 			& 166 (119 - 205)  \\
$r_m$ ($\mu$m) 		& 0.27 (0.24 - 0.35) 		& -- \\
$r_0$ ($\mu$m) 		& 4.20 (3.11 - 10.51) 		& 3.90 (3.34 - $\infty$) \\
$f$ 			& 2.5 (2.1 - $\infty$) 		& 1.34 (1.05 - 1.89) \\
\hline
\end{tabular}
\caption{Most probable values and (in brackets) 68$\%$ confidence intervals of the retrieved parameters, where $M_0$ = total mass of the plume, $r_m$ = monomer radius, $r_0$ = median particle radius, and $f$ = width of the size distribution for the small aggregate (left) and sphere-aggregate (right) particle plume solutions. The sphere-aggregate $r_0$ and the small aggregate $f$ PDFs have unconstrained upper bounds and so only the lower bounds are given, defined as the lower limit of the 68$\%$ of the accepted values immediately below the most probable value. The monomer radius for the sphere-aggregate solution is omitted, as it is multimodal and thus does not have a well-defined most-probable value or confidence interval.}
\label{tab:bestfitparams}
\end{table}
\clearpage

\appendix
\section{Sensitivity Tests}
\label{sec:appendix}

A major uncertainty in the Markov Chain Monte Carlo method is the standard deviations of the normal distributions from which the perturbations at each step are drawn. For our nominal case we use $10^4$ kg for $M_0$, 0.5 $\mu$m for $r_m$ and $r_0$, and 1 for $f$, as they are on or smaller than the value of the actual quantity itself. To test the sensitivity of our results to the standard deviations, MCMC runs are conducted where these values are halved or doubled for all parameters, while leaving all other properties of the runs unchanged. 

Figures \ref{fig:mcmcresults_frac_app} and \ref{fig:mcmcresults_sph_sph_app} show our results for the small aggregates and sphere-aggregates, respectively. In comparison with the nominal case (gray shaded regions), the halved and doubled cases are different but maintain much of the same shapes and widths. Table \ref{tab:bestfitparams_app} presents the most probable values and 68$\%$ confidence intervals of the optimized parameters for the halved and doubled cases, which are in satisfactory agreement with that of the nominal case (Table \ref{tab:bestfitparams}) to within its own 68$\%$ confidence intervals. Therefore, we can conclude that our nominal results are robust.

\setcounter{figure}{0}
\setcounter{table}{0}

\begin{figure}[p]
\centering
\includegraphics[width=0.8 \textwidth, clip=true, trim=0cm 0cm 0cm 0cm]{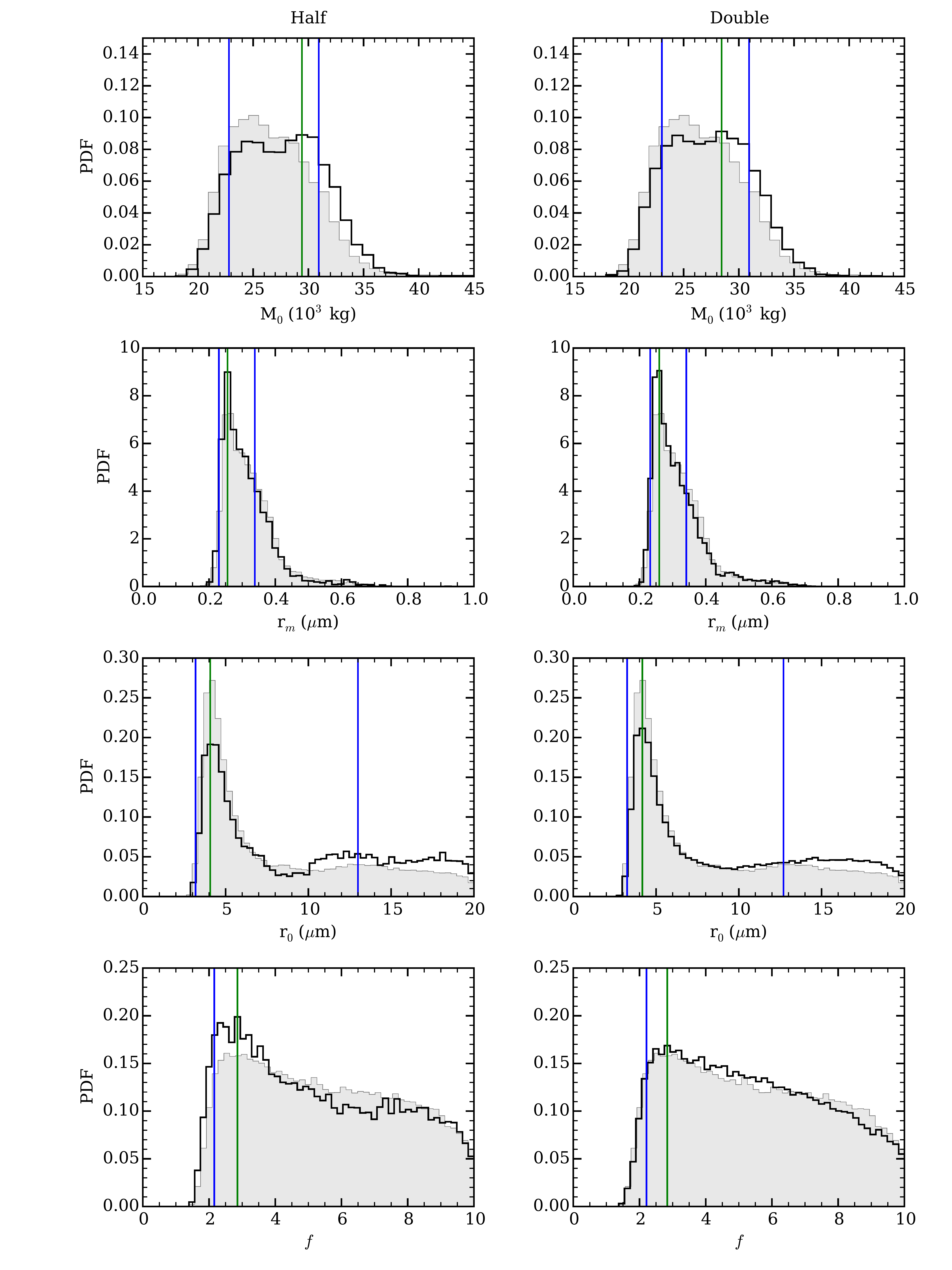}
\caption{Same as the left side of Figure \ref{fig:mcmcresults_hist} (the small aggregates), but for the sensitivity tests where we half (left) and double (right) the standard deviations of the normal distributions from which we draw the perturbations to the parameters at each step of the MCMC. The most probable values and 68$\%$ confidence intervals are given by the green and blue lines, respectively. The parameter PDFs shown on the left side of Figure \ref{fig:mcmcresults_hist} are plotted in gray for comparison.}
\label{fig:mcmcresults_frac_app}
\end{figure}
\clearpage

\begin{figure}[p]
\centering
\includegraphics[width=0.8 \textwidth, clip=true, trim=0cm 0cm 0cm 0cm]{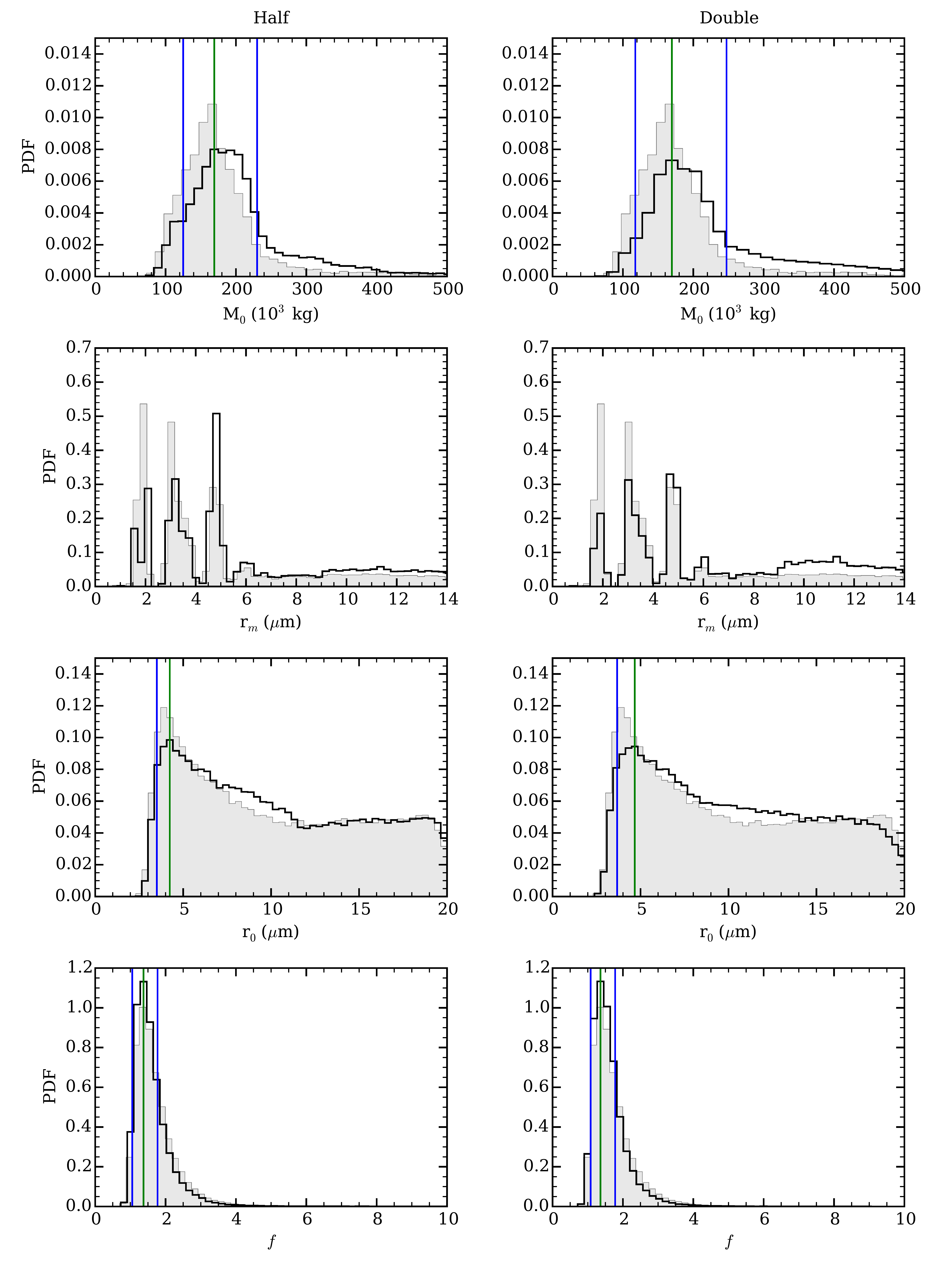}
\caption{Same as Figure \ref{fig:mcmcresults_frac_app}, but for the sphere-aggregates. The parameter PDFs shown on right side of Figure \ref{fig:mcmcresults_hist} are plotted in gray for comparison.}
\label{fig:mcmcresults_sph_sph_app}
\end{figure}
\clearpage

\begin{table}[p]
\centering
\begin{tabular}{l c c}
\hline
Parameter & Small Aggregates  & Sphere-Aggregates \\
\hline
\textit{Half:}	&		&	\\
$M_0$ (10$^3$ kg) 	& 29 (23 - 31) 			& 169 (125 - 230)  \\
$r_m$ ($\mu$m) 		& 0.26 (0.23 - 0.34) 		& -- \\
$r_0$ ($\mu$m) 		& 4.07 (3.19 - 13.00) 	 	& 4.24 (3.51 - $\infty$) \\
$f$ 			& 2.9 (2.2 - $\infty$) 		& 1.37 (1.05 - 1.77) \\
\hline
\textit{Double:}	&		&	\\
$M_0$ (10$^3$ kg) 	& 28 (23 - 31) 			& 170 (118 - 247)  \\
$r_m$ ($\mu$m) 		& 0.26 (0.23 - 0.34) 		& -- \\
$r_0$ ($\mu$m) 		& 4.17 (3.25 - 12.70) 		& 4.67 (3.67 - $\infty$) \\
$f$ 			& 2.8 (2.2 - $\infty$) 	 	& 1.36 (1.08 - 1.78) \\
\hline
\end{tabular}
\caption{Same as Table \ref{tab:bestfitparams}, but for the sensitivity tests where we half (top) and double (bottom) the standard deviations of the normal distributions from which we draw the perturbations to the parameters at each step of the MCMC.}
\label{tab:bestfitparams_app}
\end{table}
\clearpage


\end{document}